\documentclass[letter]{AA} 

\usepackage{hyperref}
\usepackage{pdflscape}
\usepackage{xcolor}
\usepackage{lscape}
\usepackage{float}
\usepackage{balance}

\usepackage{graphicx}

\usepackage{txfonts}

\begin{document}

   \title{The highly magnetic Wolf-Rayet binary \object{HD~45166} \\ resolved with VLTI/GRAVITY\thanks{Based on observations collected at the European Southern Observatory (ESO) under ESO program ID 112.25RX (PI: Shenar).}}

   \author{K.\ Deshmukh\inst{\ref{inst:kul},\ref{inst:lgi},}\thanks{Corresponding author; \texttt{kunalprashant.deshmukh@kuleuven.be}} 
          \and T.\ Shenar\inst{\ref{inst:TelAv}} 
          \and A.\ Mérand \inst{\ref{inst:eso}}
          \and H.\ Sana\inst{\ref{inst:kul},\ref{inst:lgi}}
          \and P.\ Marchant\inst{\ref{inst:kul},\ref{inst:lgi},\ref{inst:gent}}      
          \and G.A.\ Wade\inst{\ref{inst:rmcc}}  
          \and \\ J.\ Bodensteiner\inst{\ref{inst:eso},\ref{inst:api}} 
          \and A.-N.\ Chené\inst{\ref{inst:noirlab}}
          \and A.J.\ Frost\inst{\ref{inst:esoc}} 
          \and A.\ Gilkis \inst{\ref{inst:cambridge}}
          \and N.\ Langer\inst{\ref{inst:bonn},\ref{inst:mpifr}}
          \and L.\ Oskinova\inst{\ref{inst:potsdam}}
     }

   \institute{
{Institute of Astronomy, KU Leuven, Celestijnenlaan 200D, 3001 Leuven, Belgium\label{inst:kul}}
\and 
{Leuven Gravity Institute, KU Leuven, Celestijnenlaan 200D, box 2415 3001 Leuven, Belgium \label{inst:lgi}}
\and
{The School of Physics and Astronomy, Tel Aviv University, Tel Aviv 6997801, Israel\label{inst:TelAv}}
\and
{European Southern Observatory, Karl-Schwarzschild-Straße 2, 85748 Garching, Germany \label{inst:eso}}
\and
{Sterrenkundig Observatorium, Universiteit Gent, Krijgslaan 281 S9, B-9000 Gent, Belgium \label{inst:gent}}
\and
{Department of Physics and Space Science, Royal Military College of Canada, Kingston, ON, Canada K7K 7B4\label{inst:rmcc}}
\and
{University of Amsterdam, Anton Pannekoek Institute for Astronomy, Amsterdam, 1098 XH, The Netherlands\label{inst:api}}
\and
{NSF NOIRLab, 670 N. A‘ohoku Place, Hilo, HI 96720, USA\label{inst:noirlab}}
\and
{European Southern Observatory, Santiago, Chile \label{inst:esoc}}
\and
{Institute of Astronomy, University of Cambridge, Madingley Road, Cambridge CB3 0HA, UK\label{inst:cambridge}}
\and 
{Argelander Institut für Astronomie, Auf dem Hügel 71, DE-53121 Bonn, Germany \label{inst:bonn}}
\and
{Max-Planck-Institut für Radioastronomie, Auf dem Hügel 69, DE-53121 Bonn, Germany \label{inst:mpifr}}
\and
{Institute for Physics and Astronomy, Universität Potsdam, 14476 Potsdam, Germany\label{inst:potsdam}}
}

   \date{Received -; Accepted -}



\abstract{
\object{HD~45166} was recently reported to be a long-period binary comprising a B7~V star and a highly magnetic ($\langle B \rangle = 43.0\pm0.5\,$kG) hot Wolf-Rayet-like component, dubbed as a quasi Wolf-Rayet (qWR) star in literature. While originally proposed to be a short-period binary, long-term spectroscopic monitoring suggested a 22.5\,yr orbital period. With a derived dynamical mass of $2.03\pm0.44\,M_\odot$, the qWR component is the most strongly magnetized non-degenerate object ever detected and a potential magnetar progenitor. However, the long period renders the spectroscopic orbital solution and dynamical mass estimates uncertain, casting doubts on whether the qWR component is massive enough to undergo core-collapse.

Here, we spatially resolve the \object{HD~45166} binary using newly acquired interferometric data obtained with the GRAVITY instrument of the Very Large Telescope Interferometer. Due to the calibrator star being a binary as well, we implement a new approach for visibility calibration and test it thoroughly using archival GRAVITY data. The newly calibrated \object{HD~45166} data reveal the unmistakable presence of a companion to the qWR component with an angular separation of $10.9\pm0.1$\,mas (which translates to a projected physical separation of $10.8\pm0.4$\,au), consistent with the long-period orbit. We obtain a model-independent qWR mass $M_{\rm qWR} = 1.96^{+0.74}_{-0.54}\,M_\odot$ using interferometric and spectroscopic data together. This observation robustly confirms that \object{HD~45166} is truly a long-period binary, and provides an anchor point for accurate mass determination of the qWR component with further observations.
}


   \titlerunning{\object{HD~45166} resolved with VLTI/GRAVITY}
   \authorrunning{K. Deshmukh et al.}

   \maketitle
%
%
\section{Introduction}\label{sec:intro}

\object{HD~45166} is a binary system comprising a B7~V star and a hot helium star with a Wolf-Rayet (WR)-like spectral appearance \citep{vanBlerkom1978, Willis1989}. The nature of the WR component has been debated for close to a century, having been interpreted as a WR star, a Be star, and a subdwarf star \citep{Anger1933, Neubauer1948, Morgan1955, Willis1983}. Due to its relatively narrow spectral lines, peculiar abundance pattern, and spectral variability, the WR component has been dubbed a ``quasi WR star'' (qWR) since the 70s \citep{vanBlerkom1978}.  Recently, \citet{Shenar2023} detected an extremely strong magnetic field of $\langle B \rangle_{\rm qWR} = 43.0\pm0.5\,$kG in the qWR component of the system using spectropolarimetric observations secured with the ESPaDOnS spectropolarimeter of the Canada-France-Hawaii Telescope (CFHT). This detection marked the discovery of the first magnetic WR-like object known. If massive enough to undergo core-collapse, it could also be a promising progenitor of a magnetar -- a highly magnetic neutron star \citep[see][and references therein]{Kaspi2017}.

One of the most crucial uncertainties regarding \object{HD~45166} is the mass of the qWR component.  \citet{Steiner2005} reported a mass of $4.2\,M_\odot$ for the qWR component via dynamical mass measurements and adopting a 1.6\,d orbital period. However, \citet{Shenar2023} showed that the  1.6\,d period is in fact the pulsational period of the B7~V component. Relying on spectroscopic data spanning over 20\,yr, \citet{Shenar2023} found instead a long orbital period of $P = 22.5 \pm 0.5\,$yr. From a derived  evolutionary mass of $M_{\rm B} = 3.40 \pm 0.06\,M_\odot$  for the B7~V component and a mass ratio of $q = M_{\rm qWR} / M_{\rm B} = 0.60 \pm 0.13$ derived from the orbit, the authors reported $M_{\rm qWR} = 2.03 \pm 0.44\,M_\odot$ for the qWR component. 

It is uncertain whether helium stars of this mass are core-collapse progenitors \citep{2019Woosley}. The evolution models presented by \citet{Shenar2023}, which involved a merger event in what was originally a triple system, predict that the qWR component is massive enough to undergo core-collapse. This, however, cannot be claimed with certainty given the large uncertainty on its mass. Moreover, the radial-velocity (RV) amplitudes ($K_{\rm qWR} = 9.9\pm1.6\,$km\,s$^{-1}$ and $K_{\rm B} = 5.8\pm1.3\,$km\,s$^{-1}$ for the qWR and B7~V components respectively) are comparable to intrinsic variability amplitudes of both components, making the spectroscopic orbital solution, and therefore the mass ratio uncertain and challenging.

A robust confirmation that \object{HD~45166} is truly a long-period binary can be attained using long-baseline interferometry. Assuming the orbital solution derived by \citet{Shenar2023} from spectroscopy is correct, and using their derived orbital inclination of $i = 49\pm11^\circ$ from the mass calibration of the B7~V component, the projected orbital separation of the two components should be  $a\,\sin i = 10.5 \pm 1.8\,$au. At a distance of $991\pm37$\,pc measured from the {\it Gaia} parallax \citep{Bailer-Jones2021}, this translates to an angular separation of $\alpha = 11 \pm 2\,$milliarcseconds (mas). This makes \object{HD~45166} an excellent target for the GRAVITY instrument \citep{2017Gravity} of the Very Large Telescope Interferometer (VLTI), which operates in the $K$-band. The light ratio between the two components in the visual band is close to unity, implying that both components should also be easily visible in the $K$-band. The $K$-band magnitude of \object{HD~45166} ($K = 9.57$) is close to the limiting magnitude of the instrument, and requires usage of all four unit telescopes (UTs) of the VLTI.

In this letter, we present the results acquired from the first interferometric observations of \object{HD~45166} obtained with VLTI/GRAVITY. We report the detection of a companion with characteristics in agreement with the results from \citet{Shenar2023}. We demonstrate beyond doubt that \object{HD~45166} is truly a long-period binary, establishing a first interferometric data-point, and subsequently a first model-independent mass estimate for the qWR component, that can be refined with future monitoring. The letter is structured as follows: Section\,\ref{sec:obs} details the GRAVITY observations, subsequent data reduction and calibration strategies implemented for \object{HD~45166}. In Section\,\ref{sec:results}, we describe our binary search results and updated orbital parameters, followed by discussion and conclusions in Section\,\ref{sec:disc}.

\section{Observations, data reduction and calibration} \label{sec:obs}

GRAVITY is a $K$-band spectro-interferometric instrument sensitive to binary separations of $\sim$1--100\,mas and flux contrasts of $\Delta K\sim5$. \object{HD~45166} was observed with VLTI/GRAVITY in snapshot mode on 26 November 2023 (MJD 60274.33) in medium resolution mode (spectral resolving power $\lambda/\Delta\lambda = 500$) for a total integration time of 18\,min. The observation used the four UTs to form the interferometric baselines given the relative faintness of the target. To ensure that the target would be observable, we required the seeing to be smaller than 0.7'', and a DIMM seeing of $\approx 0.4''$ was indeed achieved during the exposure. 

We included a single calibrator in the observing chain (CAL-SCI). The calibrator star was \object{TYC 732-806-1}, which has a comparable $K$-band magnitude to the science target ($K = 9.49$). For data reduction and visibility calibration, we used version v1.6.0 of the standard GRAVITY pipeline \citep{2014Lapeyrere}. The observables extracted were the squared visibility (V2), closure phase (T3PHI), differential phase (DPHI) and the $K$-band spectrum (FLUX).

Unfortunately, the calibrator star \object{TYC 732-806-1} turned out to be a binary, which became obvious from its very large closure phase signal of the order of 35 degrees. We therefore fit simultaneously to the calibrator's data a binary model (two unresolved stars) as well as the transfer function (CAL-FREE approach). The transfer function consists of four closure phase offsets for the four triangles, and six affine (as function of wavelength) corrections to the squared visibility for each of the six baselines. Thankfully this approach leads to a unique solution, but with high correlation between the flux ratio of the binary and the squared visibility. To solve this, we used 32 single-star calibrators taken with the same setup on the VLTI-UTs and used the results to constrain the transfer function for \object{TYC 732-806-1} within typical ranges (CAL-PRIOR approach). Alternatively, we also applied the same self-calibration method to \object{HD~45166} itself without (SCI-FREE approach) and with prior ranges (SCI-PRIOR approach). All analyses were performed with PMOIRED\footnote{\url{https://github.com/amerand/PMOIRED}} \citep{2022SPIE12183E..1NM} version 1.2.10, which introduced the transfer function fitting. More details about this calibration method are included in Appendix\,\ref{sec:app}.

We also applied a telluric correction to the spectrum using PMOIRED, which implements molecfit \citep{2015Smette} to produce the continuum normalized spectrum (NFLUX) appropriate for further analysis.

\section{Results} \label{sec:results}

We modeled \object{HD~45166} using a parametric model in the PMOIRED software. To do so, we employed the interferometric observables V2 and T3PHI. Similar to \citet{2024Deshmukh}, we first qualitatively assessed the data to decide on a binary composed of unresolved stars (i.e. angular diameters less than approximately 0.1~mas). We decided on using V2, T3PHI and NFLUX for the modeling. Following preliminary evaluation, we performed a binary grid search (based on \citealp{2015Gallenne}) consisting of two components: (i) an unresolved star with a fixed position at origin, a flux $f_1$ as a free parameter, and emission lines seen in the spectrum; and (ii) an unresolved star with its position ($\Delta$E,$\Delta$N) varied across a grid, a flux $f_2=1-f_1$ to ensure a total continuum normalized flux of 1, and no spectral lines included. We used a grid ranging from $\Delta$E = --20 to +20\,mas and $\Delta$N = --20 to +20\,mas. The best-fit solution was obtained based on reduced chi-squared ($\chi^{2}_{\rm red}$) minimization. This method was applied for all four approaches mentioned in Section\,\ref{sec:obs}. For the final evaluation of the spectra, we used the decimated wavelets approach to fit the spectra of the two components (see \citealp{2024Sci...384..214F}).

\begin{figure*}
\sidecaption
    \includegraphics[trim={0cm 0.5cm 0cm 1.2cm},clip, width=12cm]{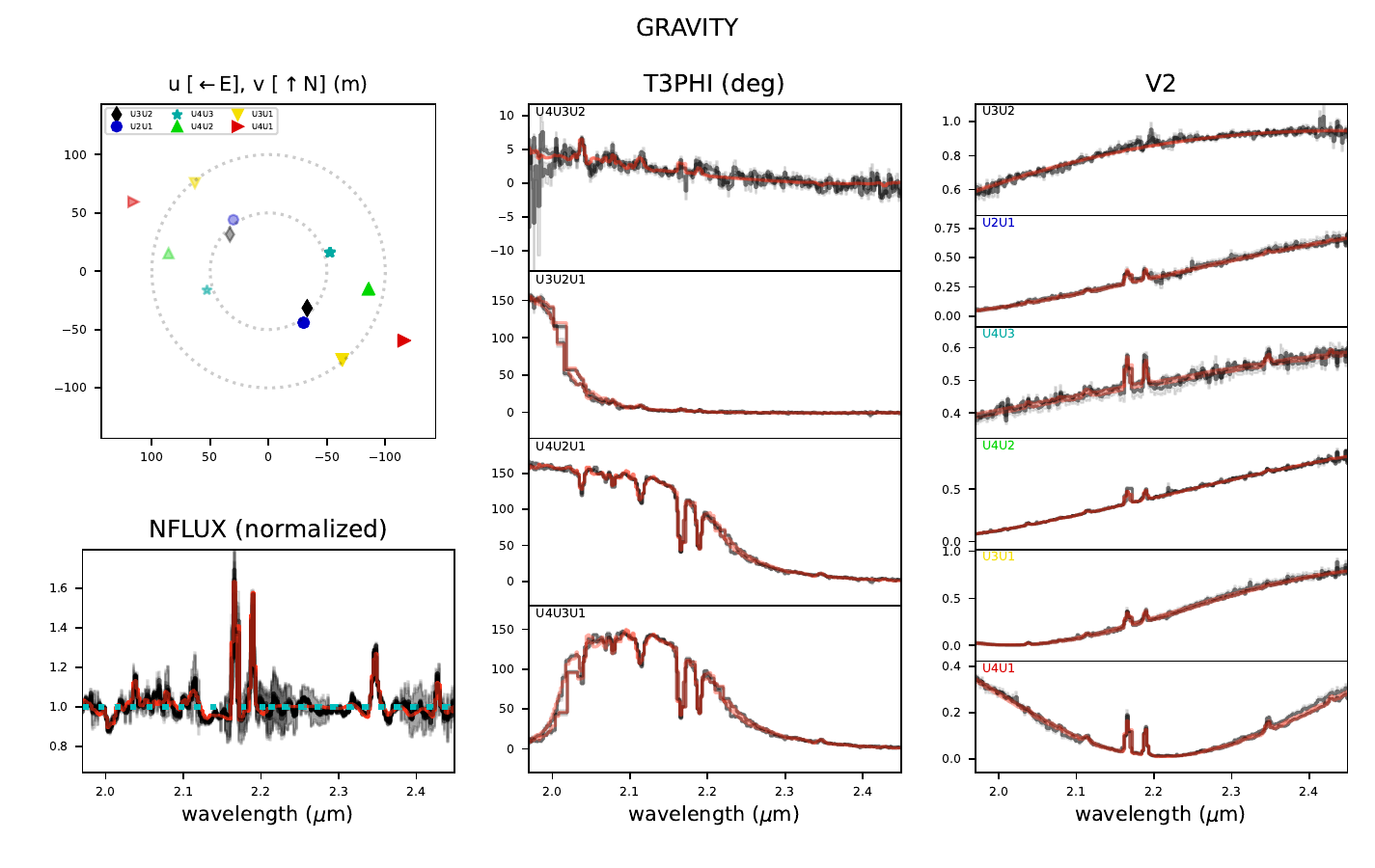}
    \caption{HD 45166 model of un-calibrated GRAVITY data, using the SCI-FREE calibration strategy (see text). The top-left panel shows the uv coverage of the observation. The bottom-left, middle and right panels show the normalized flux (NFLUX), closure phase (T3PHI) and squared visibilities (V2) respectively, with the data in black and best-fit binary model in red.}   
    \label{fig:gravity_data}
\end{figure*}

\begin{figure}
    \centering
    \includegraphics[width=1\linewidth]{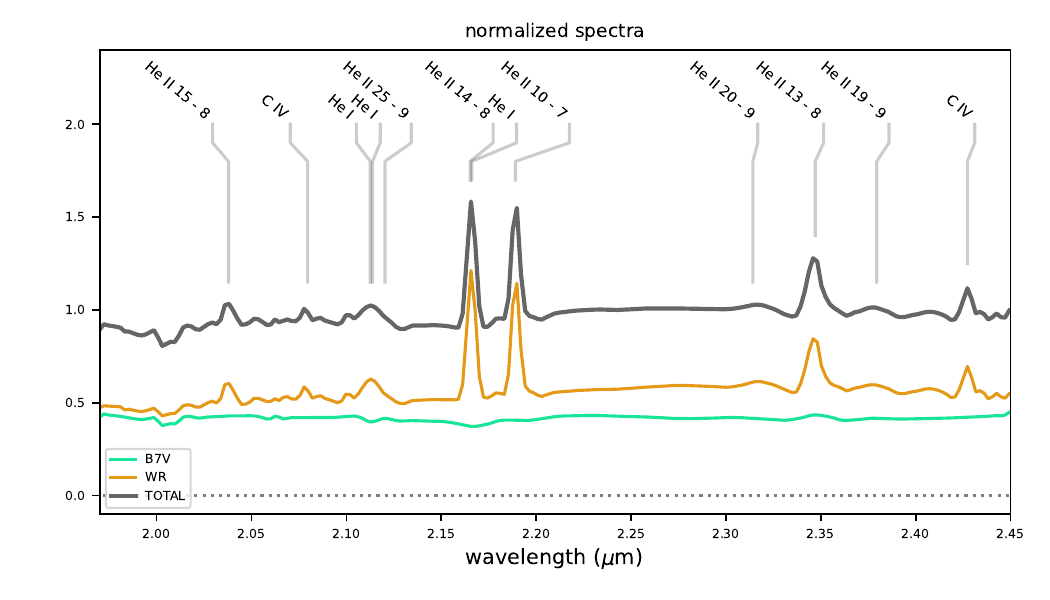}
    \caption{Derived normalized spectra in the $K$-band from the GRAVITY data, using the SCI-FREE calibration approach. The orange spectrum represents the qWR primary, the green one represents the B7~V secondary and the gray one represents their sum. Expected emission lines for the qWR are listed, based on the GRAVITY NFLUX data shown in Figure\,\ref{fig:gravity_data}.}
    \label{fig:derived_normalised_spectra}
\end{figure}

The parameters which are affected by the calibration are the continuum flux ratio and separation vector of the binary components. In our fits, these parameters had very small statistical uncertainties, much smaller than the systematics introduced by the uncertainties on the calibration. However, all four calibration strategies led to unique and very consistent binary parameters, as listed in Table\,\ref{tab:params}. As an example, Figure\,\ref{fig:gravity_data} shows GRAVITY data along with the best-fit binary model for the SCI-FREE calibration strategy. Figure\,\ref{fig:derived_normalised_spectra} shows the corresponding derived spectra for the two components, with only the qWR component possessing emission lines as expected.

\begin{table}
    \begin{center}
    \caption{\object{HD~45166} best-fit binary parameters for different calibration strategies.}
    \label{tab:params}
    \begin{tabular}{cccc}
        \hline
        \hline
         & $f_2/f_1$  & $\Delta$E (mas) & $\Delta$N (mas)\\
        Stat. uncer. & 0.003 & 0.005 & 0.008 \\
         \hline
        CAL-FREE  &  0.846 & 4.702 & 9.953 \\
        CAL-PRIOR & 0.844 & 4.867 & 9.703\\
        SCI-FREE  & 0.805 & 4.866 & 9.682\\
        SCI-PRIOR & 0.812 & 4.882 &  9.682\\
        \hline
        Adopted & $0.83\pm0.02$ & $4.83\pm0.07$ & $9.75\pm0.12$ \\
        \hline
    \end{tabular}
    \end{center}

\smallskip
\textbf{Notes:} The binary parameters listed are the flux ratio of secondary to primary $f_2/f_1$, and the secondary's position towards East and North with respect to the primary, $\Delta$E and $\Delta$N respectively in mas. For each of the parameters, we report the statistical uncertainty, the best-fit value for each calibration strategy, and the adopted (mean) value with 1$\sigma$ errors.

\end{table}

We adopted the mean values of the best-fit parameters across the four calibration strategies, resulting in a $K$-band flux ratio of the B7~V component to the qWR component $f_2/f_1 = 0.83\pm0.02$. Following the method described in Section\,5.2 of \citet{2024Deshmukh}, we obtained the absolute $K$-band magnitudes for the two components as $M_{K}^{\rm qWR} = 0.17\pm0.08$ and $M_{K}^{\rm B7~V} = 0.37\pm0.08$. The latter is consistent with a late-type B dwarf (B8~V) as per \citet{2013Pecaut}.

\begin{figure}
    \centering
    \includegraphics[trim={0cm 0.3cm 0cm 1.2cm}, clip, width=0.8\linewidth]{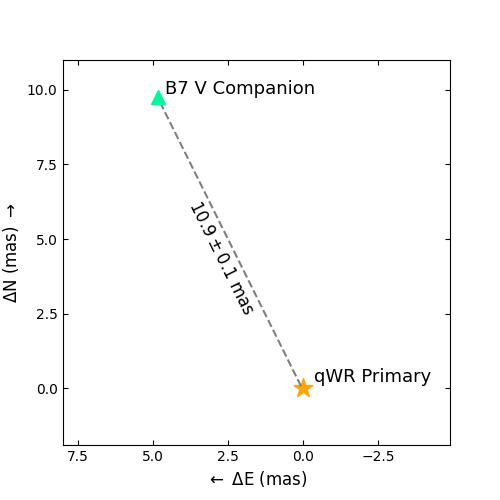}
    \caption{Relative angular positions of the two components in \object{HD~45166} with the primary fixed at origin.}
    \label{fig:relative_positions}
\end{figure}

The position of the B7~V component relative to the qWR component was found to be $\Delta$E = $4.83\pm0.07\,$mas and $\Delta$N = $9.75\pm0.12\,$mas to the East and North respectively. All errors mentioned are 68\% confidence intervals ($1\sigma$). Figure\,\ref{fig:relative_positions} illustrates the relative positions of the two components in the binary. Combining the distance to \object{HD~45166} ($991\pm37$\,pc) with the measured angular binary separation ($10.9\pm0.1\,$mas), we obtained a projected physical separation of $10.8\pm0.4\,$au.

We further computed the first 3D orbital solution for \object{HD~45166}, combining the radial velocity (RV) data from \citet{Shenar2023} and the two interferometric observables from this work ($\Delta$N, $\Delta$E, see Fig.~\ref{fig:relative_positions}). Adopting the distance to \object{HD~45166}, the new interferometric observables are sufficient to constrain the two remaining independent unknown quantities in the 3D orbital solution: the orientation in the plane of the sky ($\Omega$) and the total mass ($M_1+M_2$) or --  equivalently -- the semi-major axis of the relative orbit ($a$). To evaluate the uncertainties while taking into account the correlations between orbital parameters, we used a bootstrapping approach in which the initial data set of RVs and angular $\Delta$E, $\Delta$N positions are replaced by artificial data drawn from normal distributions centered on the observed values and with 1$\sigma$ dispersion given by the observational uncertainties. The uncertainty on the distance is also included in a similar way. We repeated the process 10,000 times and obtained the distributions of parameters shown in Figures~\ref{fig:others_corner} and \ref{fig:orbital_corner}. 

As expected, the updated orbital parameters are largely consistent with \citet{Shenar2023}, $\Omega$ being a newly constrained parameter. Thanks to interferometry, we were also able to obtain model-independent component masses which are also consistent, although with large error bars. Some relevant derived quantities include:
\begin{itemize}
\setlength\itemsep{0.5em}
    \item qWR primary mass: $M_1 = 1.96^{+0.74}_{-0.54}\,M_\odot$
    \item B7~V secondary mass: $M_2 = 3.39^{+1.00}_{-0.83}\,M_\odot$
    \item periastron distance: $a(1-e) = 7.25^{+1.57}_{-2.51}\,$au
    \item MJD for the next periastron passage: $T_{0,n} = 66325^{+490}_{-389}$, which in Besselian year is $2040.5^{+1.3}_{-1.1}$ yr
\end{itemize}

Further interferometric and spectroscopic observations will constrain these quantities to much higher precision. A more detailed description of the various orbital and derived parameters, along with figures, can be found in Appendix\,\ref{sec:corner}.

\section{Discussion and conclusions} \label{sec:disc}

We present the first interferometric observation of \object{HD~45166} using VLTI/GRAVITY, firmly establishing it as a wide binary consisting of the qWR and the B7~V components separated by $10.9\pm0.1\,$mas on the sky. This detection is a robust confirmation of \object{HD~45166} as a long-period binary system. We also obtain the first dynamical mass measurement for the qWR, validating assumptions made in \citet{Shenar2023} and therefore their conclusions.

The measured distance at periastron of $\sim7\pm2\,$au provides strong evidence that the system is not a product of a previous mass transfer phase from the qWR to its B7 V companion. In that case, the qWR would have been the product of binary stripping rather than a merger as suggested by \citet{Shenar2023}. To produce a $2\,M_\odot$ helium core requires a $\sim 8\,M_\odot$ progenitor, which would not fill its own Roche lobe before core-helium depletion at a periastron distance of 7 au. Although the current separation would be different from that at the onset of mass transfer, this does not impact our conclusion. Given our derived mass ratio $q=1.7\pm0.2$ and the one at the onset of the hypothetical mass transfer phase, $q_\mathrm{i}\sim 3.39\,M_\odot/8\,M_\odot=0.42$ (assuming non-conservative mass transfer), the pre-interaction semi-major axis would have been similar to that currently observed. This would exclude binary stripping and instead favor the merger scenario of \citet{Shenar2023}. Further observations are critical to completely exclude solutions with a much smaller periastron distance (see Fig\,\ref{fig:others_corner}).

Within the uncertainties on the mass of the qWR star the resulting outcome could either be the formation of an ONeMg white dwarf or to undergo an explosion, possibly through an electron-capture supernova \citep{2022Chanlaridis}. Additionally, in binary systems stripped stars can undergo an additional phase of mass transfer after core-helium depletion which at short orbital periods ($\lesssim$ 1 day) is expected to remove most of the helium rich layers and produce ultra-stripped supernovae \citep{2015Tauris}. Such a scenario was recently suggested for the formation of the neutron star in the high mass X-ray binary CPD-29 2176 \citep{2023Richardson}. For the case of the qWR star, however, the newly confirmed wide orbit implies a separation too large for future binary interaction and its final fate will be determined by its own evolution, with further constraints on its mass being critical to determine the actual outcome.

As part of our analysis, we adopted a modified calibration strategy for GRAVITY data, motivated by a bad calibrator star. TYC 732-806-1, the calibrator star also turned out to be a binary, rendering it unsuitable for visibility calibration using traditional methods. We implemented a novel method to bypass the issue as described in Section\,\ref{sec:obs}, which can also be applied to other GRAVITY data affected by bad calibrator stars.

In addition to confirming the binary status of \object{HD~45166}, our GRAVITY observation serves as the first relative astrometric measurement of the binary. Given its estimated orbital period of 22.5\,yr, follow-up GRAVITY observations over the coming years will provide more astrometric measurements, eventually enabling the determination of a visual orbit for the binary. Combining it with spectroscopic observations will enable us to determine much more precise model-independent masses for both components of the binary. This work therefore represents a crucial step toward determining the mass of the qWR component in \object{HD~45166}, and subsequently its true nature as a potential magnetar progenitor.

\section*{Data availability}

All GRAVITY data used in this letter are available on the ESO Archive. Reduced data may be made available upon request to the authors.

\begin{acknowledgements}
 Based on observations made with ESO Telescopes at the La Silla Paranal Observatory under programme ID 112.25RX.001. Based on data obtained from the ESO Science Archive Facility under request number 936605. This research has used measurements obtained at the Mercator Observatory which receives funding from the Research Foundation – Flanders (FWO) (grant agreement I000325N and I000521N). KD and HS acknowledge funding from grant METH/24/012 at KU Leuven. TS acknowledges support from the Israel Science Foundation (ISF) under grant number 0603225041. GAW acknowledges the Discovery Grant support from the Natural Sciences and Engineering Research Council (NSERC) of Canada. PM acknowledges support from the FWO senior fellowship number 12ZY523N. 

\end{acknowledgements}

\bibliographystyle{aa}
\bibliography{main}

\clearpage

\begin{appendix}

\section{Calibration of the GRAVITY data}
\label{sec:app}

Interferometric data suffer from instrumental and atmospheric effects which result in a multiplicative bias for the visibility amplitude, and additive biases for the phase closure. The usual procedure to account for these effects consists of observing a partially resolved star as an on-sky calibrator to limit the uncertainty introduced in the process (see for example \cite{2005A&A...433.1155M}). This calibrator needs to be taken close in time (less than 1 hour) within the science target observation, and close-by on sky with the same observing model.

In the case of the observations presented in this work, the calibrator unfortunately turned out to be a binary with a strong signal in both closure phase and visibility. Before deciding to use valuable telescope time and redo the observations with a different calibrator, we tested the idea that the binary signal could be disentangled from the transfer function (TF) signal. This turned out to work remarkably well as we describe in this appendix.

\subsection*{Validation on archival single-star calibrators}

The first step is to check what are the historical values and scatter of the TF for data taken with the same mode and with the same telescopes. For the closure phase, we chose a TF in the form of an additive bias (dominated by birefringence effect in the beam combiner): $\mathrm{T3PHI_{true} = T3PHI_{obs}} + bias$. For the visibility squared amplitude, we chose a multiplicative factor $a$ with a slope $b$ in wavelength (dominated by atmospheric turbulence, which has a strong chromatic dependency): $\mathrm{V^2_{true}} = a\mathrm{V^2_{obs}}[1+b(\lambda-2.2\,\mu\mathrm{m})]$. The binarity, on the other hand, manifests as a strong chromatic signal which we verified \textit{a posteriori} can be disentangled from the TF signal. 

\begin{figure*}
    \centering
    \includegraphics[scale=0.8]{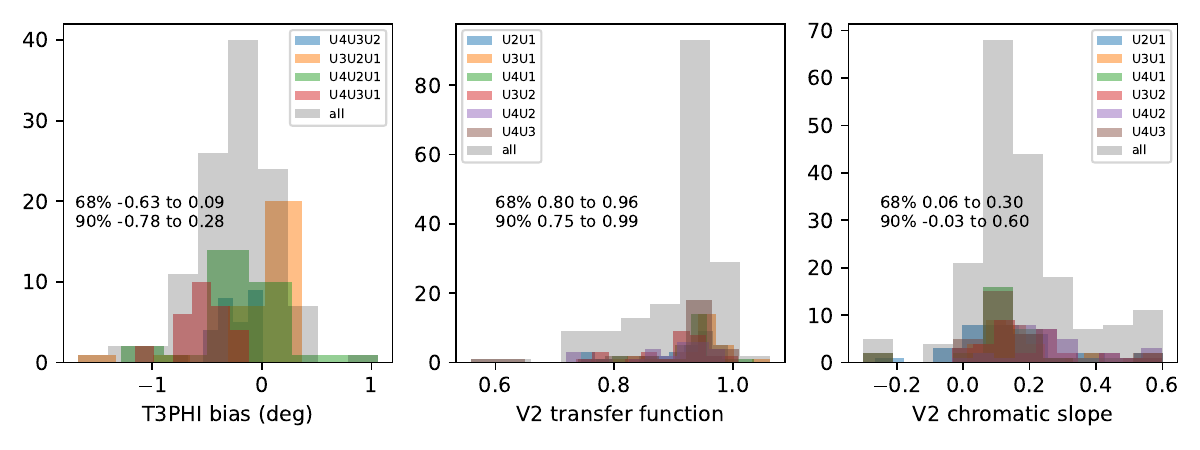}
    \caption{Histograms of calibration parameters fitted to historical good calibrators (partially resolved stars) taken with the same mode and telescopes in the year prior to our data. The three panels cover the 3 types of parameters. Left: closure phase bias in degrees; center: V2 multiplicative factor; right: V2 chromatic slope}
    \label{fig:calibrators_UTs}
\end{figure*}

We requested from the ESO archive all the proper calibrator observations taken the year prior to our observation in the same mode (medium spectral resolution, combined polarizations) and with the same telescopes (VLTI-UTs), resulting in 32 individual observations\footnote{Available as ESO archive request number 936605}. We applied our calibration strategy by fitting the transfer function and assuming the angular diameters from the JMMC Stellar Diameters Catalog \citep{2017yCat.2346....0B}. The historical calibrators turned out to have tight scatter of the transfer function parameters. For the closure phase, the bias is typically close to 0, and less than 1 degree in absolute value (see Fig.~\ref{fig:calibrators_UTs}). For the squared visibilities, 90\% of the data have a multiplicative factor $a$ between 0.75 and 0.99, and a chromatic slope $b$ between $-0.03$ and 0.6. 

\subsection*{Coping with a binary as a calibrator}

We first perform a binary grid search using as starting point the most probable TF values of $bias$=0 for all triangles and $(a,b)=(0.95,0.2)$ for all baselines as initial values. In this process, 19 parameters are fitted: all except the sizes of the stars in the binary which are set to 0.05\,mas based on surface brightness relations. 

The binary solution with free TF parameters has a $\chi^2_{\rm red}$=1.56 (with 1839 degrees of freedom) and the following transfer function values: closure phase biases are all between 0.7 and 8.8 degrees, $a$ between 0.61 and 0.95, and $b$ between 0.11 and 0.62, relatively close to historical values, especially for the V2 parameters (Fig.~\ref{fig:calibrators_UTs}). 

In the fitting process, we can add priors to the transfer function parameters based on our analysis of the historical transfer function values: 
\begin{itemize}
\item the closure phase transfer function biases are set to 0$\pm$1 degrees
\item the multiplicative transfer function $a$ of the squared visibilities are assumed to be 0.9$\pm$0.3
\item the chromatic slope $b$ of the squared visibilities are assumed to be 0.3$\pm$0.4
\end{itemize}

With priors, the binary solution has a $\chi^2_{\rm red}$=2.32  and very reasonable transfer function values: closure phase biases are all between $-0.04$ and +0.06 degrees, $a$ between 0.82 and 0.90, and $b$ between 0.25 and 0.42, very much inline with historical values (Fig.~\ref{fig:calibrators_UTs}). 

As a check, using a single-star model and no priors on TF values results in a $\chi^2_{\rm red}$=7.4 and transfer function values widely incompatible with the historical values, by dozens of sigmas. If we force a prior on the TF, we get $\chi^2_{\rm red}$=326, which for degrees of freedom (d.o.f.) close to 2000 is a clear indication the single-star model is incompatible with the data. All results are summarized in table~\ref{tab:bin_calibrator_models}.

\subsection*{Calibration of HD~45166}

Based on the convincing binary fit of the calibrator, we applied the same strategy to HD~45166 by fitting a binary model and the transfer function parameters. CAL-FREE corresponds to fix the TF parameters to the ones computed from the calibrator (first line from Table~\ref{tab:bin_calibrator_models});  CAL-PRIOR uses the TF parameters determined on the calibrator, but using priors based on historical values (second line from Table~\ref{tab:bin_calibrator_models}). Conversely, SCI-FREE and SCI-PRIOR are self-calibration approaches where the TF parameters are fitted simultaneously to the binary parameters, with or without priors. The fact that all of these four calibration approaches lead to similar solutions is a strong indication that this calibration method is appropriate in this case.

We cannot guarantee that this approach would work for any observation of any object. In the case of our HD~45166 observations, our calibration seems to work primarily because the TF and astrophysical signals from our model (unresolved binary stars with free spectra) are very distinct, which can hence be determined independently from the minimization algorithm. For example, the closure phase TF is assumed to be a small (of the order of 1 degree or less) bias, where the astrophysical signal has a large amplitude (up to 150 degrees) and strong chromatic variations. Regarding the visibility squared amplitudes, we can note that baselines U3U3 and U3U1 reach maxima (nearly 1), whereas baseline U4U1 reaches minimum (nearly 0) which again is very distinct from the TF signal (see Fig.~\ref{fig:gravity_data}).

\begin{table*}
    \begin{center}
    \caption{Various models applied to the calibrator TYC~732-806-1.}
    \label{tab:bin_calibrator_models}
    \begin{tabular}{ccccccc}
        \hline
        \hline
        Model & TF prior & T3PHI $bias$ & V$^2$ $a$ & V$^2$ $b$ & $\chi^2_{\rm red}$ \\
        & & 90\%: -0.8 to 0.3 & 90\%: 0.75 to 0.99 & 90\%: 0.0 to 0.6 & d.o.f.=1839\\ 
        \hline 
        binary & no & 0.7 to 8.8 & 0.61 to 0.95 & 0.11 to 0.62 & 1.56\\
        binary & yes & -0.04 to 0.06 & 0.82 to 0.90 & 0.25 to 0.42 & 2.32\\
        single & no & -25 to 0 & 0.22 to 0.86 & 0.0 to 0.57 & 7.14 \\
        single & yes & -7.3 to 0 & 0.22 to 0.86 & 0.20 to 0.37 & 326\\
        \hline
    \end{tabular}
    \end{center}
    
\medskip
\textbf{Notes:} Shown are the ranges of fitted transfer function parameters (one for each triangle or baseline) depending of the model (binary or single star), and whether or not priors were used (based on historical data, shown in the second line).

\end{table*}

\section{Additional plots}
\label{sec:corner}

\begin{figure*}
\sidecaption
    \includegraphics[width=12cm]{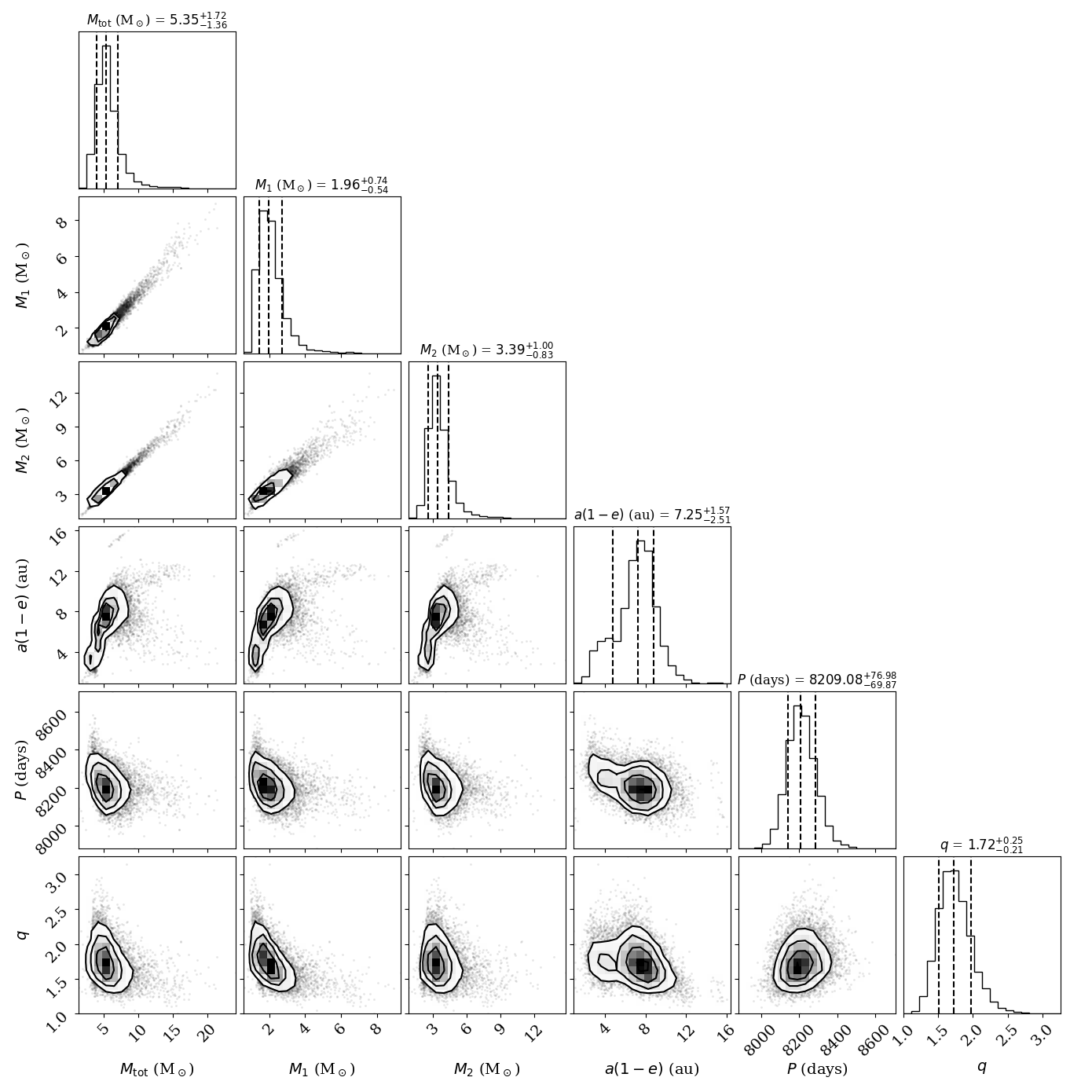}
    \caption{Corner plot showing key parameters for \object{HD~45166} and their 1$\sigma$ confidence intervals obtained from the bootstrapping analysis, namely: total mass of the binary ($M_{\rm tot}$), mass of the qWR primary ($M_1$), mass of the B7~V secondary ($M_2$), periastron distance ($a(1-e)$), orbital period ($P$), and mass ratio of the secondary to primary ($q$).}   
    \label{fig:others_corner}
\end{figure*}
\begin{figure*}
\sidecaption
    \includegraphics[width=12cm]{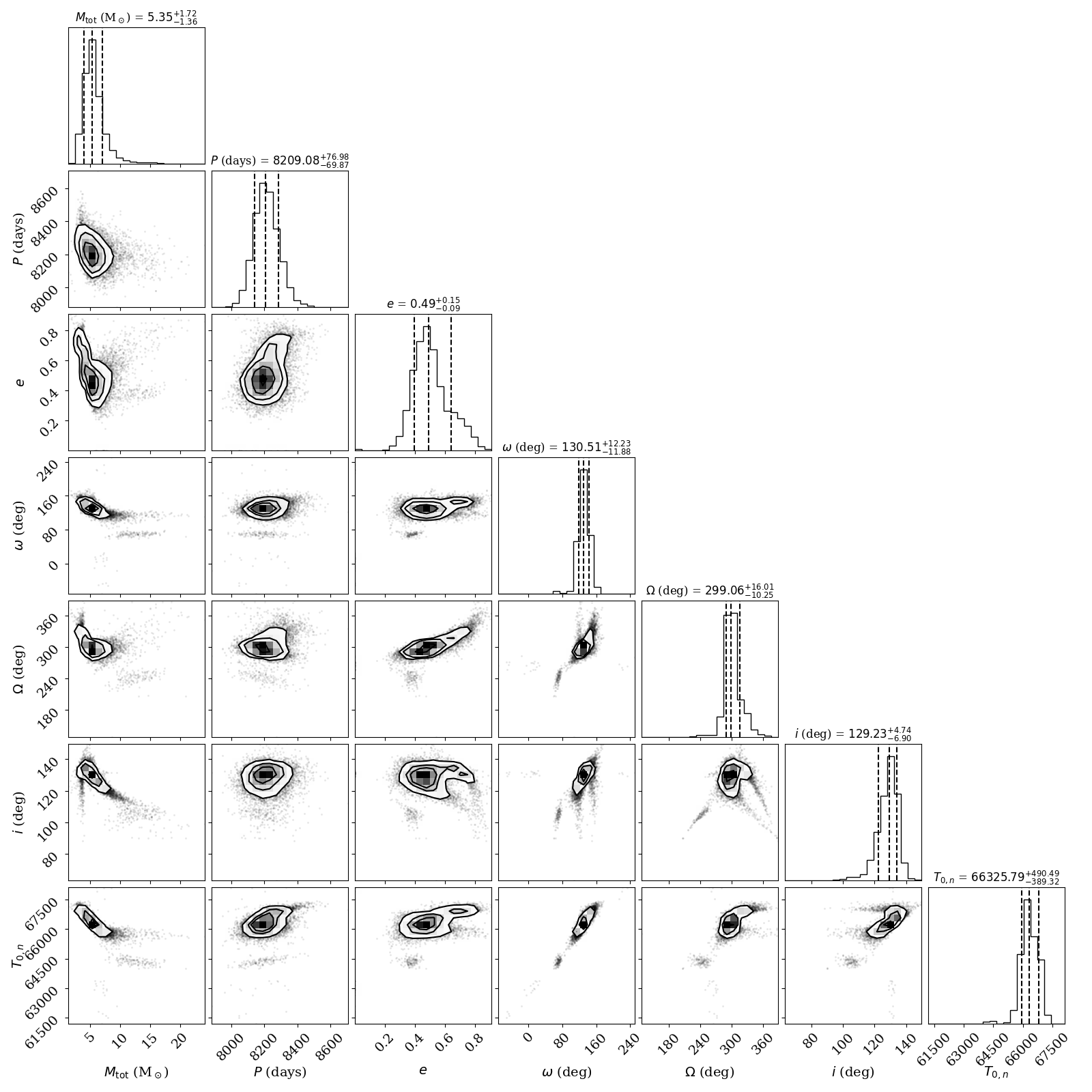}
    \caption{Corner plot showing updated orbital parameters for \object{HD~45166} and their 1$\sigma$ confidence intervals obtained from the bootstrapping analysis, namely: total mass of the binary ($M_{\rm tot}$), orbital period ($P$), eccentricity ($e$), argument of periapsis ($\omega$), longitude of ascending node ($\Omega$), inclination ($i$) and MJD for the next periastron passage ($T_{0,n}$).}   
    \label{fig:orbital_corner}
\end{figure*}

Using the method described in Section\,\ref{sec:results}, we derived six important parameters for \object{HD~45166}: total mass of the binary ($M_{\rm tot}$), mass of the qWR primary ($M_1$), mass of the B7~V secondary ($M_2$), periastron distance ($a(1-e)$), orbital period ($P$), and mass ratio of the secondary to primary ($q$). Figure\,\ref{fig:others_corner} shows the corner plot for all six parameters. We also show the usual orbital parameters along with the total mass of the binary from our bootstrapping run in Figure\,\ref{fig:orbital_corner} for completeness. The next (`n') periastron passage ($T_{0,n} = 66326^{+490}_{-389}$) is expected in about 15 years from now. We also calculated the next times when true anomaly ($\vartheta$) equals $\vartheta_1=-\omega$ or $\vartheta_2=\pi-\omega$ representing orbital phases in which the projected separation equals actual separation ($T_{1,n} = 65334^{+634}_{-360},T_{2,n} = 66695^{+293}_{-318}$). Nevertheless, the interferometric orbit can be refined significantly with follow-up observations in the next few years.

In contrast to \citet{Shenar2023}, the component masses derived here are model-independent. However, they have significant uncertainties introduced by the combination of uncertainties on the total mass and the mass ratio. While further epochs will ensure much higher precision on the total mass as well as other orbital parameters, the precision on component masses will still be limited by the precision on mass ratio, or equivalently, on RV semi-amplitudes. This also motivates continued spectroscopic follow-up of \object{HD~45166}.

\end{appendix}

\end{document}